# Greening our Laws:
# Revising Land Acquisition Law for Coal Mining in India

By Sugandha Srivastav* and Tanmay Singh**

*British Academy Postdoctoral Fellow, Smith School of Enterprise and the Environment, University of Oxford. E: sugandha.srivastav@smithschool.ox.ac.uk.

** Senior Litigation Counsel at Internet Freedom Foundation. E: tanmay.pratap@gmail.com

**Abstract**

Laws that govern land acquisition can lock in old paradigms. We study one such case: the Coal Bearing Areas Act of 1957 (CBAA) which provides minimal social and environmental safeguards, and deviates in important ways from the Right to Fair Compensation and Transparency in Land Acquisition, Rehabilitation and Resettlement Act, 2013 (LARR). The lack of due diligence in the CBAA confers an undue comparative advantage to coal development, which is inconsistent with India's current stance to phasedown coal use, reduce air pollution, and advance modern sources of energy. We argue that the premise under which the CBAA was historically justified is no longer valid due to significant changes in the local context. Namely, the environmental and social costs of coal-based energy are far more salient and the market has cleaner energy alternatives that are cost competitive. We recommend updating land acquisition laws to bring coal under the general purview of LARR or, at minimum, amending CBAA to ensure adequate environmental and social safeguards are in place, both in letter and practice.

## 1. Introduction

The coal industry in India saw its genesis in the 18th century. Three centuries of growth have led to the formation of institutional structures designed to support its development. This includes access to loans at preferential rates[1], import bans to shield against international competition, and lenient time extensions in installing mandatory air pollution control equipment (Garg, Viswanathan, Narayanaswamy and Ganesan 2019). However, amongst the numerous examples of preferential treatment that the coal sector has received, one of the most overlooked ones is coal mining projects' easy access to land, which reduces the cost of new coal development and significantly overlooks societal safeguards.

Land acquisition for coal mining is conducted under the Coal Bearing Areas Act of 1957 (CBAA), which allows the Government of India to exercise significant levels of discretion when acquiring land. The law is historically justified on grounds of India's national interest. Land acquisition for general infrastructure projects is conducted under the Right to Fair Compensation and Transparency in Land Acquisition, Rehabilitation and Resettlement Act, 2013 (LARR) which has provisions for ensuring social and environmental justice.

India's "managed approach" to the coal sector is nested in a view that coal is a key strategic asset (Lahiri-Dutt 2014a). The CBAA's text states that this law: "establish[es] greater public control over the coal mining industry and its development…" (CBAA 1957, § 9A). However, the factual condition which made coal a strategic asset in the 20th century is questionable today. Due to the presence of cost competitive alternative energy resources and an understanding of the negative health and environmental impacts of coal combustion, it is harder to assert that CBAA, with its sweeping levels of discretion and its lack of environmental and social safeguards, acts in the national interest. The challenge is best summarised by Lahiri-Dutt (2016), who asserts: "Coal India Limited (CIL) has not, in its lifetime, met the expectations of a socially and environmentally responsible corporation – such as due diligence in land acquisition, resettlement of displaced people, rehabilitation of the environment, and financial viability." Recent attempts by CIL to engage in building small-scale community infrastructure have been an attempt to rebuild social legitimacy and distract from the systemic ways in which its operations have caused the displacement and marginalization of communities (Herbert and Lahiri-Dutt 2014).

---

[1] For example, when the sector was under financial duress in the 1990s, the World Bank bailed it out by giving loans at highly subsidised rates.

Coal has exceptionally high environmental and social externalities that include climate change, air pollution, destruction of natural habitats, displacement of indigenous communities, water table depletion, biodiversity loss, etc. (European Commission 2009). Studies show that the negative externalities of coal are so large that they often exceed its market price per ton (Cardoso 2015). Communities close to coal mines in India suffer from significantly higher levels of child mortality (Barrows, Garg and Jha 2018). Moreover, once coal's negative health effects are taken into account, then the savings from shutting down coal mining operations can be several times greater than the cost of installing new renewable energy and battery storage infrastructure to replace lost capacity (Rafaty, Srivastav and Hoops 2020). Air pollution, which is a by-product of coal combustion, increases disease burden, lowers life expectancy and results in premature deaths (Balakrishnan et al 2019). It also reduces labour productivity (Hanna and Oliva 2011; Graff-Zivin and Neidell 2012). None of these numerous negative externalities are reflected in the price of coal.

## 2. Land acquisition for coal overlooks social protections

The CBAA outlines the process by which the government can acquire private lands for coal mining. The government is first required to issue a notification stating its interest to prospect for coal on private land (CBAA 1957, § 4.1). Upon issuing the notification, the government can begin surveying the land; digging into the sub-soil; demarcating boundaries; felling crop, fence or jungle that is in the way; and undertake all other acts necessary for prospecting the land. The only safeguard is that the government cannot enter enclosed buildings and living spaces without giving seven days of notice in writing. There is no provision to ensure such notifications actually reach the affected communities, many of which live in rural areas without access to channels of information that typically convey such information. The burden of communication is often carried out by NGOs who travel to affected rural communities to convey that there land is under risk of being acquired.[2]

Within two to three years, the government can declare its power to acquire rights over the land, and separately declare acquisition, if coal is discovered. At this stage, the landowner may raise objections which must be heard and responded to by the competent authority. If the government is satisfied after hearing such objections, it can award compensation to the affected landowner and take possession of the land (CBAA 1957, § 12 and 13). If the government believes it is necessary to acquire the land immediately, for *any* reason, it may dispense with the need for hearing of objections all together (CBAA 1957, § 9A).

---

[2] Based on authors' interview with climate journalist, Aruna Chandrasekhar.

The LARR which governs land acquisition for general development projects has significantly more protections enshrined within it (see Table 1). For example, while the LARR requires a social impact assessment (SIA) and a resettlement & rehabilitation plan for displaced persons, the CBAA requires neither. The exemptions to the LARR are specified in Section 105 under the Fourth Schedule (in addition to coal, 12 other land acquisition acts are exempted including those that pertain to railways, atomic energy, ancient monuments, petroleum, national highways, and electricity).

**Table 1**

|  | **LARR** | **CBAA** |
|---|---|---|
| SOCIAL IMPACT ASSESSMENT | Required | Not required |
| PUBLIC HEARINGS/CONSULTATIONS WITH COMMUNITY LAND OWNERS | Required | Not required |
| INDEPENDENT APPRAISAL OF SOCIAL IMPACT | Required | Not required |
| PUBLICATION OF INTENTION TO PROSPECT | In Official Gazette, two daily local newspapers, at local government offices in local language, uploaded on government website | In Official Gazette only. |
| ADVANCE WARNING FOR PROSPECTING | The time taken for preparation of the social impact assessment serves as advance notice | None |
| COMPENSATION | Required | Required |
| REHABILITATION AND RESETTLEMENT SCHEME | Required | Not required |
| PROTECTION FOR VULNERABLE CASTES AND TRIBES | Acquisition to be avoided, as far as possible. Last resort to be demonstrated. | None |
| PROTECTION FOR PREVIOUSLY DISPLACED PERSONS | Not to be displaced, or double compensation to be paid | None |
| EXPEDITED PROCEDURES | Urgency condition limited by defence, natural calamity or has to go to parliament for other | Urgency condition with no definition of what counts as urgency |

Under the LARR, the government is required to prepare a SIA in consultation with local administration before acquiring the land (LARR 2013, § 4.1). The SIA must include an estimate of the affected families; the extent of lands, houses and other common properties likely to be affected; whether the land proposed for acquisition is the bare minimum required for the stated purpose; whether alternative land has been considered and found not to be feasible; and an analysis of the social costs of the projects versus potential benefits (LARR 2013, § 4.4). Under the CBAA by contrast, there is no requirement for a SIA in any form.

Similarly, the government is required, under the LARR, to prepare a resettlement and rehabilitation scheme which contains details about the compensation, replacement house, resettlement area, land allotments, subsistence allowance, transport allowance, and mandatory employment for affected families (LARR 1957, § 31). Only after this has been completed, does the government have the power to take possession of the lands (LARR 2013, § 38). Under the CBAA, there is no mandate for devising a rehabilitation and resettlement scheme. The only form of compensation is monetary, which includes the market value of the land, damage to any standing crops or trees, damage to any immoveable property on the land, and reasonable expenses incidental to changing place of residence or business of the affected landowner (CBAA 1957, § 13.5). However, the market value of the land is, in practice, decided at the date at which the government declares its intent to acquire the land rather than the date of actual acquisition. Sometimes the gap between the two can be years, even a decade, during which time, the market value of the land has significantly appreciated. Moreover, accounts of misleading local communities through false promises of mining jobs are extensively documented in audit reports (Herbert and Lahiri-Dutt 2004, Herbert and Lahiri-Dutt 2014).

A third layer of protection – public hearings – is also denied to indigenous communities when coal is involved. In the LARR, a public hearing is required before the SIA to ascertain and record the views of the affected families (LARR 2013, § 5). The report is published and evaluated by an independent multidisciplinary expert group (LARR 2013, § 7.1 and 7.2). This expert group must consist of two non-government social scientists, two representatives of the local administration; two experts on rehabilitation; and a technical expert in the subject relating to the purpose for land acquisition. If the expert group finds that the project does not serve a public purpose, or that the social costs outweigh the economic benefits, then no further steps can be taken by the government to acquire the land (LARR 2013, § 7.4). The government may overrule the recommendation of this expert group, but it will have to record its reasons for doing so in writing. Any person objecting to the acquisition of their land must be provided an opportunity to be heard, and the government's decision in respect of such objections must be recorded in writing (LARR 2013, § 15). In the CBAA, there is no requirement for public hearings or independent assessments.

In the CBAA, communities only have 30 days from the moment the government declares it power to acquire the land to raise any objections. Since there is no obligation on the government to ensure communities reliably receive information (which the LARR has, via public hearings), windows of opportunities to raise objections often go missed. In practice, many communities in coal bearing areas do not have access to the Official Gazette in which the government publishes notifications and therefore get no advance warning. The costs of ensuring information reaches affected communities often falls on third-parties such as NGOs that go into affected areas to communicate the information contained in gazettes.

The government may also take immediate possession of the land and abbreviate the land acquisition process as necessary, pursuant to an urgency provision in the LARR (LARR 2013, § 40.1). However, the government's power to invoke this urgency provision is restricted to the minimum area of land required for specified reasons of national security, that is for the defence of India or for emergencies arising out of natural calamities (LARR 2013, § 40.2). The urgency provision under the CBAA, by contrast, is not limited in any way. In "cases of urgency", defined as moments when the "government is satisfied that it is necessary to acquire immediately the whole or any part of the land", the right to object to the acquisition for any reason is revoked. Given that there are no restrictions around what constitutes an "urgency" in the CBAA, the level of discretionary power given to the government is immense (Ramanathan 2011). This is also evidenced by the numerous times the Supreme Court of India has had to uphold the right of a landowner to object to the acquisition of their land in the absence of a real urgency.[3]

The CBAA's focus towards compulsory acquisition therefore constitutes effectively unchallengeable value judgements: if the government wishes to mine coal, there are no social costs which will stand to outweigh the assumed economic benefits and stop the project from advancing.

---

[3] Laxman Lal v. State of Rajasthan, (2013) 3 Supreme Court Cases 764, at paragraph 22.4; Anand Singh v. State of Uttar Pradesh, (2010) 11 Supreme Court Cases 242, at paragraph 42; Darshan Lal Nagpal v. Government of NCT of Delhi, (2012) 2 Supreme Court Cases 327, at paragraph 36.

## 3. Coal mining expansions are exempt from Environmental Impact Assessments

Under the 2006 EIA Notification, an environmental impact assessment (EIA) is required for coal mining but a series of executive orders have significantly reduced the thrust of this regulatory requirement. For example, in December 2012, under pressure from the Ministry of Coal, the Environment Ministry exempted coal projects from the requirement to get additional clearances and hold public consultations to expand operations by 25%. In January 2014, the Environment Ministry exempted any expansion of an existing project, having up to 8 million tonnes per annum (MTPA) capacity, by up to 50%. In July 2014, 8 MTPA was increased to 16 MTPA, and in September 2014 to 20 MTPA. Finally in July 2017, coal mining received a blanket exemption from conducting any public hearing for capacity expansions of up to 40% (Expert Appraisal Committee, 2017).[4]

An example of a similar regulatory laxity is related to the Environment Ministry's 2015 standards for limiting SO2, NOx and other emissions. Adhering to the new regulatory requirements necessitates coal-fired powerplants to install pollution control equipment such as scrubbers. However, compliance rates remain very low, and the deadline for compliance has been extended to 2022 and likely to be extended again (Garg, Viswanathan, Narayanaswamy and Ganesan 2019).

## 4. Roadmap for change

The CBAA provides considerably lower protections to indigenous communities and affected persons as compared to the LARR, and the issue of whether a separate pathway for coal mining is necessary is ripe for constitutional consideration.

The factual condition which justified coal's preferential access to land via the 1957 CBAA rested on arguments regarding national interest and energy security. In the early 20th century, there was truth to these claims: coal was the dominant source of energy and cost-competitive domestic alternatives were not readily available. Even though coal mining came with negative consequences such as local air pollution, community displacement and the destruction of traditional economic livelihoods, these outcomes were tolerated in the greater interest of nation building. The transformation of Jharkhand (literally meaning, *the land of forests*) to a landscape of gaping holes caused by rampant open pit mining was deemed to be a net positive.

---

[4] Minutes of the 15th meeting of the Expert Appraisal Committee dated July 25, 2017, Agenda 15.5 at page 22. Available at http://environmentclearance.nic.in/writereaddata/Form-1A/Minutes/010820176ABWO9WXApprovedMOM15thEACheldon25July2017Coal.pdf (Last accessed on August 15, 2021.

However, today coal has a variety of competitors which include, but are not limited to, solar and wind energy. Economic analysis has shown that only coal burnt at the pithead is cost competitive with solar energy (Wood Mackenzie 2019; Tongia and Gross 2019). In most cases, when coal is transported, renewable power has a lower levelized cost (Tongia and Gross 2019). Estimates show that by 2040, India can meet 80% of its energy demand through wind and solar energy at a cost that is at least as competitive, if not more, than a coal-based energy trajectory (Lu et al 2020). The CBAA is not consistent with India's commitment to phasedown coal use by 2070.

Legislators must therefore re-visit laws pertaining to coal to make sure they reflect the context of today: the exceptional leniency accorded by the CBAA must be re-examined and due diligence must be the new focus to ensure appropriate environmental and social protections. The CBAA as it stands today acts like an implicit subsidy to coal development.

Progressive policy would seek to involve coal mining within the remit of LARR. The due diligence procedures under LARR which include social impact assessments, rehabilitation, and mandatory public hearings, are necessary to ensure that coal mining is conducted responsibly. There are two possible routes to align India's legal regime with its climate objectives. First, the Indian legislature must bring the land acquisition process for coal mining under the LARR, and repeal the CBAA. This will require parliamentary amendments. Second, judicial intervention by constitutional courts under India's writ jurisdiction is necessary to provide greater rights-based protections against coal pollution. Both routes require sustained advocacy and public participation.

We conclude that identifying and amending anachronisms in law is an important part of facilitating the energy transition, which is needed to mitigate anthropogenic climate change and deleterious air pollution. Judicial and parliamentary interventions can help align India's legal regime closer to its climate needs for the future.